\documentclass[amsmath, amssymb, aps, pra,twocolumn,floatfix,nofootinbib]{revtex4}
\usepackage[utf8]{inputenc}
\usepackage{titlesec}
\usepackage[colorlinks]{hyperref}
\usepackage{amsmath,amssymb}
\usepackage{bbold}
\usepackage{float}
\usepackage{dcolumn}
\usepackage{natbib}
\usepackage{bm}
\usepackage{dsfont}
\usepackage{url}
\usepackage{mathtools}
\usepackage{amsthm}
\usepackage{amsfonts}
\usepackage{braket}
\usepackage{cancel}
\usepackage{bbold}
\usepackage{makecell}
\usepackage[normalem]{ulem}
\usepackage{xcolor}
\usepackage{comment}

\usepackage{afterpage}

\begin{document}
\title{Converting entanglement into ensemble basis-free coherence}
\author{A.\,Kodukhov$^{*}$}
\affiliation{\text{Terra Quantum AG, Kornhausstrasse 25, St. Gallen, Switzerland, 9000}
\\$^*\,$ak@terraquantum.swiss
}
\begin{abstract}
The resource theory of coherence addresses the extent to which quantum properties are present in a given quantum system.
While coherence has been extensively studied for individual quantum states, measures of coherence for ensembles of quantum states remain an area of active research.
The entanglement-based approach to ensemble coherence---arising from the measurement--ensemble duality principle and the Born rule---connects the ensemble coherence with both the entanglement resource and the measurement's uncertainty.
This paper presents two methods for generating ensemble coherence from a fixed amount of entanglement between two qubit systems.
The first method involves applying a von Neumann measurement to one part of a non-maximally entangled bipartite state, resulting in a pair of non-orthogonal states whose coherence can equal the initial entanglement.
The second method considers a class of symmetric observables capable of generating ensembles used in quantum key distribution (QKD) protocols such as B92, BB84, and three-state QKD.
As a result, this work contributes to understanding how much ensemble coherence can be obtained from a given amount of entanglement.
\end{abstract}

\maketitle
\section{Introduction}
Since the establishment of quantum mechanics, there has been an ongoing debate about the boundary between quantum and classical phenomena.
Bohr and Einstein famously disagreed over the existence and implications of the uncertainty principle as a genuinely quantum feature.
This foundational debate ultimately favored Bohr’s interpretation~\cite{jammer1974philosophy}.
In the second half of the 20th century, attention shifted toward understanding whether it is possible to extract partial information about two non-commuting observables, particularly in the context of the double-slit interference experiment~\cite{wootters1979complementarity, englert1996fringe}.

Today, we are witnessing the rapid emergence of quantum technologies, promising breakthroughs in computation and secure communication.
These developments require a formal and quantitative understanding of the quantum properties of the states employed in quantum devices.
In the modern framework of quantum information theory, this role is fulfilled by the resource theory of coherence.
The resource theory of coherence has been thoroughly developed for individual density matrices~\cite{baumgratz2014quantifying, luo2017PRA, qi2018quantifying, yuan2017quantum}, with~detailed analyses of its dynamics under various decoherence processes relevant to quantum algorithms, metrology and other areas~\cite{streltsov2017colloquium, shlyakhov2018quantum}.
Coherence is also studied in the conditions of relativistic motion that appeared to be an additional source of decoherence~\cite{relativistic_2024}.
Importantly, resource theories also provide a structured framework for converting different quantum quantities---for example, determining how much coherence can be generated from a given amount of entanglement, and~vice~versa~\cite{piani2011all, streltsov2015measuring, ma2016converting, chitambar2019quantum, takahashi2022creating}.

Although the relative entropy of coherence is well studied for single quantum states, it has not yet been fully developed for ensembles of quantum states.
The main interest of the current research is to quantitatively characterize the quantum properties of an ensemble and to determine under which conditions a system can be regarded as quantum.
This problem naturally arises in another important area of quantum technologies, quantum cryptography, where it is necessary to characterize the properties of ensembles of bit-encoding quantum states.
Legitimate users face a tradeoff: the states of the ensemble must be sufficiently indistinguishable to prevent an eavesdropper from gaining information about the key, yet sufficiently distinguishable for Bob to reliably obtain information about each bit~\cite{kodukhov2023boosting}.
This challenge motivates the extension of the resource theory of coherence to ensembles of quantum~states.

The active discussion of how much quantumness is contained in an ensemble of states was initiated by C.A.\,Fuchs in the context of two non-orthogonal states~\cite{fuchs2002just}.
As a measure of ensemble quantumness, he proposed, for~example, the~gap between the Holevo quantity~\cite{holevo} and the accessible information~\cite{divincenzo2004locking, konig_renner_2007, pastushenko2023improving}, as~well as the achievable fidelity in cloning an unknown state drawn from the ensemble.
Later, C.A.\,Fuchs and M.\,Sasaki introduced a related quantumness measure based on the worst-case difficulty of transmitting ensemble states through a classical communication channel~\cite{fuchs2003quantumness}.
Subsequent studies have proposed a variety of alternative measures for ensemble quantumness.
In Refs.~\cite{sun2021quantumness, sun2021post, fan2024quantifying}, the~measure is based on the Gram matrix of the ensemble.
This matrix, which can be interpreted as a density matrix, encodes the pairwise overlaps between all states in the ensemble.
Quantumness is then quantified using the relative entropy of coherence~\mbox{\cite{gour2009measuring, baumgratz2014quantifying, winter2016operational}} or the $l_1$-norm~\cite{baumgratz2014quantifying, shao2015fidelity} of the Gram matrix.
Another approach defines ensemble quantumness via the Rényi entropy of the Gram matrix~\cite{Renyi_2022}.
In Ref.~\cite{qi2018quantifying}, a~different measure is introduced based on unitary similarity-invariant norms of the commutators between the ensemble's density matrices.
A final, but~equally important, approach defines ensemble quantumness as the average (over the ensemble) of the relative entropy of coherence, minimized over all possible bases.
This quantity was first introduced in Refs.~\mbox{\cite{luo2009relative, luo_How_quantum, luo2011quantumness}} and was independently developed further in Refs.~\cite{kronberg2019coherence, kodukhov2023entanglement}.
Importantly, this measure is basis-free and effectively quantifies the extent to which the states of an ensemble cannot be simultaneously~diagonalized.

Ref.~\cite{kodukhov2023entanglement} analytically demonstrated that an ensemble can be generated by performing a local measurement on one part of an entangled bipartite state~\cite{hughston1993complete}.
When an observer applies an arbitrary POVM to their subsystem, an~ensemble of post-measurement states is created in the other subsystem.
The resulting ensemble coherence (or quantumness) depends on both the entanglement consumption and the uncertainty of the applied measurement, where the latter corresponds to the unpredictability of measurement outcomes.
The original motivation for connecting ensemble coherence with the measurement of an entangled state comes from the duality principle, which maps a POVM to an ensemble and vice~versa~\cite{hall1997quantum, dall2011informational, holevo2012information}, and~from the conservation of Born-rule probabilities under the duality mapping.
The main goal of this study is to integrate the quantum measurement formalism, the~Born rule within the duality principle, and~the existing resource theory in order to quantitatively characterize the quantum properties of ensembles of quantum~states.

The results of this study demonstrate a strategy for generating coherence by applying a class of rank-1 symmetric observables to qubit systems.
For a specific range of entanglement, the~method is optimal in the sense that the entire initial entanglement is converted into coherence, saturating the upper bound derived in Ref.~\cite{kodukhov2023entanglement}.
This class of symmetric observables allows for the generation of ensembles corresponding to various QKD protocols, including B92~\cite{bennett1992quantum}, BB84~\cite{bennett2014}, and~three-state QKD~\cite{phoenix2000three, boileau2005unconditional}.
This connection aligns with the main interest of characterizing quantumness in the context of quantum cryptography.
Overall, the~study helps establish how the amount of achievable ensemble coherence depends on the available entanglement~resource.

\bigskip
\section{Entanglement-based approach to coherence}
In this section, we describe the entanglement-based approach to quantum ensemble coherence and define crucial quantum information characteristics, such as the relative entropy of coherence of a single state, ensemble basis-free coherence, and~the uncertainty quantity of a~measurement.

An ensemble of quantum states can be obtained by a local measurement of one subsystem of a bipartite entangled state.
Let Alice\,(A)---the holder of one of two finite-dimension subsystems---apply a local positive operator-valued measure\,(POVM) to her part of the entangled state.
Then, the~holder of the second subsystem, Bob\,(B), observes an ensemble of quantum states.
The observed density matrix on Bob's side depends on Alice's obtained measurement result.
The process of ensemble generation was initially suggested in Refs.
~\cite{hughston1993complete, bennett1996concentrating} and fully formalized in Ref.~\cite{kodukhov2023entanglement}.

The entangled state can be written in the Schmidt decomposition
\begin{equation}
    \ket{\Psi}_\text{AB} = \sum\limits_{n=1}^d \sqrt{\lambda_n} \ket{n}_\text{A} \ket{n}_\text{B},
\end{equation}
where $d$ is the dimension of each subsystem, $\sqrt{\lambda_n} $ are Schmidt coefficients and $\{\ket{n}_\text{A}\ket{n}_\text{B}\}_n$ is the Schmidt basis.
In the Schmidt basis, the~partial density matrices of each subsystem naturally has the diagonal form
\begin{equation}
    \sigma_\text{A(B)} = \sum\limits_{n=1}^d\lambda_n \, \ket{n}\!\bra{n}_\text{A(B)}.
\end{equation}
A
 pure bipartite entangled state can be characterized by its entanglement measure. 
In the simplest case, the~entanglement is defined as a the von Neumann entropy of any of two subsystems~\cite{bennett1996concentrating, bennett1996purification, wootters1998entanglement}
\begin{equation}
    E\left( \ket{\Psi}_\text{AB} \right) = S(\sigma_\text{B}) = -\sum\limits_{n=1}^d \lambda_n \log_2 \lambda_n.
\end{equation}
The latter equality arises when considering the diagonal form of the reduced density matrices.
In this case, the~entanglement corresponds to the classical Shannon entropy of the squared Schmidt coefficients.
The shared state between Alice and Bob is maximally entangled when all Schmidt coefficients are equal.
Conversely, the~state is separable when only one Schmidt coefficient is~non-zero.

Alice applies observable $\mathcal{M}_\text{A}$ with the POVM elements $M_i^{(\text{A})}$ ($i=1,\dots,N$) to her part of the entangled state.
Bob observes an ensemble $\mathcal{E}_\text{B}$ of quantum states $\rho_i$ with probabilities $p_i$
\begin{equation}
    \mathcal{E}_\text{B}\!=\!\!\left\lbrace p_i \!= \text{tr} \left(\sigma_\text{A}M_i^{(\text{A})} \!\right),\,\rho_i \!= \frac{1}{p_i} \sqrt{\sigma_\text{B}} M_i^{(\text{B})} \sqrt{\sigma_\text{B}} \right\rbrace_i,
    \label{eq:ensemble}
\end{equation}
where
\begin{equation}
    M_i^{(\text{B})} = \sum\limits_{m,n} \bra{m}M_i^{(\text{A})}\ket{n}_\text{A} \ket{n}\!\bra{m}_\text{B}.
\end{equation}
The ensemble obeys the similar probability distribution as measurement results on Alice's side.
Operators $M_i^{(\text{B})}$ act in Bob's~subsystem.

The original motivation for generating ensembles through entangled-state measurements comes from the concept of duality.
To each observable, one can associate a dual ensemble, and~vice~versa~\cite{hall1997quantum, dall2011informational, holevo2012information}.
The joint probability distribution of measuring a state from an ensemble and obtaining a particular measurement outcome, given by the Born rule, is preserved when the ensemble and observable are mapped to their dual counterparts, a~POVM and an ensemble, respectively.
Equation~(\ref{eq:ensemble}) coincides with the expression dictated by the duality concept.
Thus, the~entanglement-based approach naturally arises from duality and the conservation of Born-rule probabilities under the duality~transformation.

Alice's POVM can be characterized by the quality of uncertainty with respect to the measured entangled state.
Motivation for this quality is raised from the entropic uncertainty relations (EURs) for a single observable~\cite{krishna2002entropic, coles2017entropic, hall2018entropic}.
Firstly, we define the entropy of measurement results, when a POVM $\mathcal{M}$ is applied to an arbitrary density matrix
\begin{equation}
    U\left( \mathcal{M}|\rho \right) = -\sum\limits_{i=1}^N \text{tr}\left( M_i \rho \right) \log_2 \text{tr}\left( M_i \rho \right),
\end{equation}
where $M_i$ is a POVM element of the considered observable, $\rho$ is a probe density matrix, and $\text{tr}\left( M_i \rho \right)$ is the Born-rule probability.
The uncertainty of POVM with respect to the bipartite entangled state shows the unpredictability of the measurement results, when a POVM is applied to one of the subsystems
\begin{equation}
    \begin{gathered}
        U\left( \mathcal{M}|\ket{\Psi}_\text{AB} \right) = \min_{\mathcal{E}_\text{A}} \sum\limits_k p(k) \, U\!\left( \mathcal{M}|\rho_k \right),\\
        \mathcal{E}_\text{A}=\{ p(k), \rho_k \}_k : \sum\limits_k p(k) \rho_k = \sigma_\text{A}.
    \end{gathered}
\end{equation}
The minimum is taken over all possible decompositions of Alice's reduced density matrix $\sigma_\text{A}$, and~is achieved by decompositions consisting of pure~states.

The scenario in which Alice generates an ensemble by applying a POVM to an entangled state is closely related to the framework of quantum memory-assisted entropic uncertainty relations (EURs)~\cite{berta2010uncertainty, coles2017entropic, maroulakos2025majorana}.
In this framework, Alice and Bob share a bipartite state, which may be entangled.
Alice measures one of two observables and then announces her choice to Bob.
Bob, who keeps his part of the state in quantum memory, tries to guess Alice's measurement outcome.
The uncertainty relation for the two chosen observables depends on both their complementarity and the entanglement of the initial Alice–Bob state. Certain variations in quantum memory EURs are also sensitive to entanglement consumption~\cite{tomamichel2012framework}, i.e.,~the difference between the initial entanglement resource and the residual entanglement.
A key difference between the quantum memory EUR framework and the entanglement-based approach to coherence is that the latter considers the uncertainty of a single publicly known POVM, rather than two~observables.

Before providing the coherence definition for an ensemble, we remind the framework for a single quantum state.
To define coherence of a density matrix $\rho$, one usually utilizes the relative entropy of coherence~\cite{baumgratz2014quantifying, streltsov2017colloquium}
\begin{equation}
    C_T^{\text{rel}}(\rho) = S(\rho|\rho^\text{diag}) = S(\rho^\text{diag})-S(\rho),
    \label{eq:rel_ent}
\end{equation}
where $S(\rho| \sigma)=\text{tr}\rho(\log_2\rho-\log_2\sigma)$ is the relative entropy~\cite{vedral2002role} and $\rho^\text{diag}$ is the diagonal version of the considered density matrix with respect to some fixed orthonormal basis $T=\{ \ket{e_k} \}_{k=1}^d$:
\begin{equation}
    \rho^\text{diag} = \sum\limits_{k=1}^d \bra{e_k} \rho \ket{e_k} \ket{e_k}\!\bra{e_k}.
\end{equation}
The choice of the basis $T$ can be described by a unitary transformation that maps the original computational basis to a specific one:
\begin{equation}
    T = \sum_{k=1}^d \ket{e_k}\!\bra{k}:\quad \forall k \,\,\ket{e_k} = T \ket{k}.
\end{equation}
The relative entropy of coherence can be averaged over an ensemble in a reference basis $T$:
\begin{equation}
    C_T (\mathcal{E}) = \sum\limits_i p_i C_T^{\text{rel}}\left(T \rho_i T^\dag \right),
\end{equation}
where $\mathcal{E}=\lbrace p_i, \rho_i \rbrace_i$ is a considered ensemble and $T \rho_i T^\dag$ is the form of the density matrix $\rho_i$ in the basis $T$.

Here, we study the basis-free coherence~\cite{kronberg2019coherence} of an ensemble, that is obtained from the latter quantity as a minimum over all orthonormal bases:
\begin{equation}
    C(\mathcal{E}) = \min_T C_T(\mathcal{E}) = \min_T \sum\limits_i p_i C_T^{\text{rel}}\left(T \rho_i T^\dag \right).
    \label{eq:ensemble_coherence}
\end{equation}
A similar definition of the ensemble's basis-free coherence is introduced in Refs.~\cite{luo2009relative, luo_How_quantum, luo2011quantumness}.
The described definition of ensemble basis-free coherence is an extension of the already developed framework of the relative entropy of coherence for a single quantum state, discussed in detail in Ref.~\cite{streltsov2017colloquium}.
Basis-free ensemble coherence shows to what extent density matrices of an ensemble can be simultaneously diagonalized in a common~basis.

Choosing a basis means that Bob performs a von Neumann measurement on his subsystem in order to infer Alice’s outcome.
The Born rule specifies the probability of each possible measurement result.
The probability distribution of outcomes is usually sufficient to determine informational quantities such as mutual information or error probability.
However, the~Born rule does not describe how the quantum state is modified after measurement.
In the entanglement-based framework for ensemble coherence, knowledge of the joint probability distribution of Alice’s and Bob’s measurement results alone is not sufficient to determine ensemble coherence.
For computing the first term in the relative entropy of coherence (Equation~(\ref{eq:rel_ent})), the~joint distribution is enough.
For the second term, however, the~exact form of Bob’s ensemble is required.
Thus, in~addition to the Born rule, the~state-update rule is also necessary~\cite{shrapnel2018updating}, which in our case follows from the duality principle.
Although the Born rule does not specify Alice’s post-measurement state, this is irrelevant for our purposes, since we are not concerned with the post-measurement state of her subsystem.
Bob’s resulting ensemble depends only on the original POVM, and~not on the underlying quantum instrument~\cite{davies1970operational}.
Consequently, the~ensemble is sensitive to the process of information extraction from Alice’s system, but~not to its physical~modification.

The studied ensemble coherence is, in~fact, a generalization of the standard relative entropy of coherence for a single state (Equation~(\ref{eq:rel_ent})).
One can always reduce ensemble coherence to single state coherence by shrinking an ensemble into a trivial ensemble consisting of a single state.
Consider an arbitrary ensemble $ \{ p_i, \eta_i \}_i\,(i=1,\dots ,N)$ and a fixed basis $T$.
Then, we add an arbitrary density matrix $\rho$ as a perturbation in the examined ensemble.
Let $\delta$ be the probability of the
perturbation density matrix in the modified ensemble $\mathcal{E}$:
\begin{equation}
    \mathcal{E} = \left\lbrace \left( \delta, \rho \right) \left( (1-\delta)p_1, \eta_1\right),..., \left( (1-\delta)p_N, \eta_N \right)  \right\rbrace.
\end{equation}
When the perturbation parameter is zero, $\delta = 0$, we have the original ensemble.
With the increase in $\delta$, the ensemble loses its initial properties and starts to behave as a single state $\rho$.
As a result, when the perturbation parameter approaches 1, the~ensemble coherence in the basis $T$ is equal to the relative entropy of coherence of $\rho$:
\begin{equation}
\lim \limits_{\delta  \rightarrow 1} C_T(\mathcal{E})
      = C_T^{\text{rel}} (\rho).
\end{equation}
Thus, the~definition of ensemble coherence possesses the continuity property~\cite{winter2016operational, winter2016tight} and can be considered a general quality of~quantumness.

In the previous study~\cite{kodukhov2023entanglement}, the~main boundaries on the basis-free coherence were established.
Firstly, ensemble's coherence is always upper-bounded by the Holevo quantity~\cite{holevo} of the ensemble:
\begin{equation}
    C\left( \mathcal{E} \right) \leq \chi \left( \mathcal{E} \right) = S \left( \sum\limits_{i=1}^N p_i \rho_i \right) - \sum\limits_{i=1}^N p_i S\left( \rho_i \right).
    \label{eq:upper}
\end{equation}
The right-hand side is closely related to the entanglement consumption.
The first term corresponds to the entanglement of the initial Alice--Bob state, while the second term is equal to the average entanglement remaining after the measurement.
If the applied measurement is a rank-1 POVM, the~ensemble consists only of pure states, and~thus, its Holevo quantity is equal to the initial amount of entanglement~resource.

In addition, ensemble's coherence is lower-bounded by the gap between the Holevo quantity and accessible information of the ensemble
\begin{equation}
    C\left( \mathcal{E} \right) \geq \chi \left( \mathcal{E} \right)-I_\text{acc} \left( \mathcal{E} \right)
\end{equation}
This lower bound was independently presented in Refs.
~\cite{luo2009relative, luo_How_quantum, luo2011quantumness,kodukhov2023entanglement}.
The Holevo quantity characterizes the maximum amount of information that can be extracted from an ensemble using collective measurements, whereas the accessible information refers to the maximum information obtainable via individual measurements~\cite{shor2002number, konig_renner_2007, suzuki2007accessible, pastushenko2023improving}.
The presence of quantum properties in an ensemble is indicated by the advantage of collective measurements in terms of extractable information.
If all states in the ensemble can be diagonalized in the same basis, the~Holevo quantity is achieved by the von Neumann measurement in that basis.
In such cases, the~information extractable via collective measurements coincides with the accessible~information.

\bigskip
\section{Pair of non-orthogonal states}\label{sec:B92}
In this section, we show the first example of obtaining coherence from entanglement.
The example consists of the von Neumann measurement of a part of an entangled~state.

We consider the qubit case ($d = 2$) and a von Neumann measurement in the Hadamard basis, $\mathcal{M}_\text{H} = \{ \ket{+}\!\bra{+}, \ket{-}\!\bra{-} \}$, applied to Alice's subsystem.
The Hadamard basis is mutually unbiased with respect to the computational basis, which forms part of the Schmidt basis.
Despite the simplicity of the considered POVM, it generates an ensemble with non-zero coherence.
Let Alice and Bob share a non-ideally entangled bipartite state:
\begin{equation}
\begin{gathered}
    \ket{\Psi}_\text{AB} = \cos \alpha \ket{0}_\text{A}\ket{0}_\text{B}+\sin \alpha \ket{1}_\text{A}\ket{1}_\text{B},\\
    0 \leq \alpha \leq \pi/2,
\end{gathered}
    \label{ent_state}
\end{equation}
where $\cos \alpha$ and $\sin \alpha$ are the Schmidt coefficients.
For $\alpha = \pi / 4$, we have the maximally entangled state.
Applying the observable $\mathcal{M}_\text{H}$ to Alice's subsystem produces the ensemble of two pure states $\mathcal{E}_\text{B92}=\{ p_i, \ket{\psi_i}_\text{B}  \}_{i=1}^2$ in the subsystem B.
Both measurement results are equiprobable; thus, the~states in the ensemble are also equiprobable ($p_1=p_2=1/2$).
The states in Bob's subsystem have the following form:
\begin{equation}
    \begin{gathered}
        \ket{\psi_1} = \cos \alpha \ket{0}_\text{B}+\sin \alpha \ket{1}_\text{B},\\
        \ket{\psi_2} = \cos \alpha \ket{0}_\text{B}-\sin \alpha \ket{1}_\text{B}.
    \end{gathered}
    \label{Nonorthogonal_ensemble}
\end{equation}
The obtained ensemble corresponds to the B92 QKD protocol~\cite{bennett1992quantum} where two non-orthogonal states are utilized to encode logical bits ``0'' and ``1''.
In B92 QKD, Alice transmits encoded states through a quantum channel to Bob, who measures the incoming states using unambiguous state discrimination (USD); see, e.g.,~in Ref.~\cite{filippov2020operational}.
Bob's POVM consists of three elements:
$M_1=q(\mathds{1} - \ket{\psi_2}\!\bra{\psi_2})$, $M_2=q(\mathds{1} - \ket{\psi_1}\!\bra{\psi_1})$, and $M_?=\mathds{1}-M_1-M_2$, where $q = 1/(1+|\braket{\psi_1|\psi_2}|)$.
The first two POVM elements correspond to logical bits, while “?” denotes inconclusive measurement outcomes that are discarded during postselection.
In the absence of an eavesdropper, Bob makes no errors.
The probability that a transmitted state passes postselection is $p_\checkmark = 1-|\braket{\psi_1|\psi_2}|$.
A crucial characteristic of any QKD protocol, the~key generation rate, is proportional to the informational advantage of Alice and Bob over the eavesdropper~\cite{devetak2005distillation}.

In the coherence framework, if~the initial Alice--Bob state is maximally entangled\linebreak ($\alpha = \pi /4$), the~resulting ensemble is a pair of orthogonal states.
For $\alpha =0$ (zero entanglement), both ensemble's states are similar and coincide with the basis state $\ket{0}_\text{B}$.
In the mentioned cases, no significant quantum properties are expected, as~the states in the ensemble can be simultaneously diagonalized.
In a potential QKD scenario, these situations lead to a zero key generation rate.
When the states in the ensemble are orthogonal, an~eavesdropper can measure all transmitted states without being detected and obtain full information about them.
Thus, legitimate users cannot achieve any informational advantage over the eavesdropper.
When the states are identical, Bob, applying the USD measurement, always obtains inconclusive outcomes and therefore cannot distinguish between logical bits.
Both scenarios result in the absence of secret key generation, which parallels the absence of ensemble~coherence.

For other entanglement values, states of the ensemble are non-orthogonal and, thus, cannot be diagonalized simultaneously.
Coherence of an ensemble in a fixed reference basis is the entropy of a probability distribution appearing during the von Neumann measurement of an ensemble in this basis.
To calculate the coherence, we fix an arbitrary orthogonal basis $T = \{ \ket{e_1}, \ket{e_2} \}$ in the subsystem B:
\begin{equation}
    C_T(\mathcal{E}_\text{B92}) = \frac{1}{2} h_2\left(|\braket{e_1|\psi_1}|^2\right) + \frac{1}{2} h_2\left(|\braket{e_1|\psi_2}|^2\right),
    \label{pair_in_basis}
\end{equation}
where $\ket{e_1} = \cos \theta_T \ket{0}_\text{B}+\sin \theta_T \ket{1}_\text{B}$, $\theta_T$ is a real-value basis parameter and $h_2(x)=-x\log_2x-(1-x)\log_2(1-x)$ is the binary entropy.
Then, we need to minimize the latter equation over all bases.
We have only one real-value parameter in the optimization~problem.

\begin{figure}[H]
\includegraphics[width=9cm]{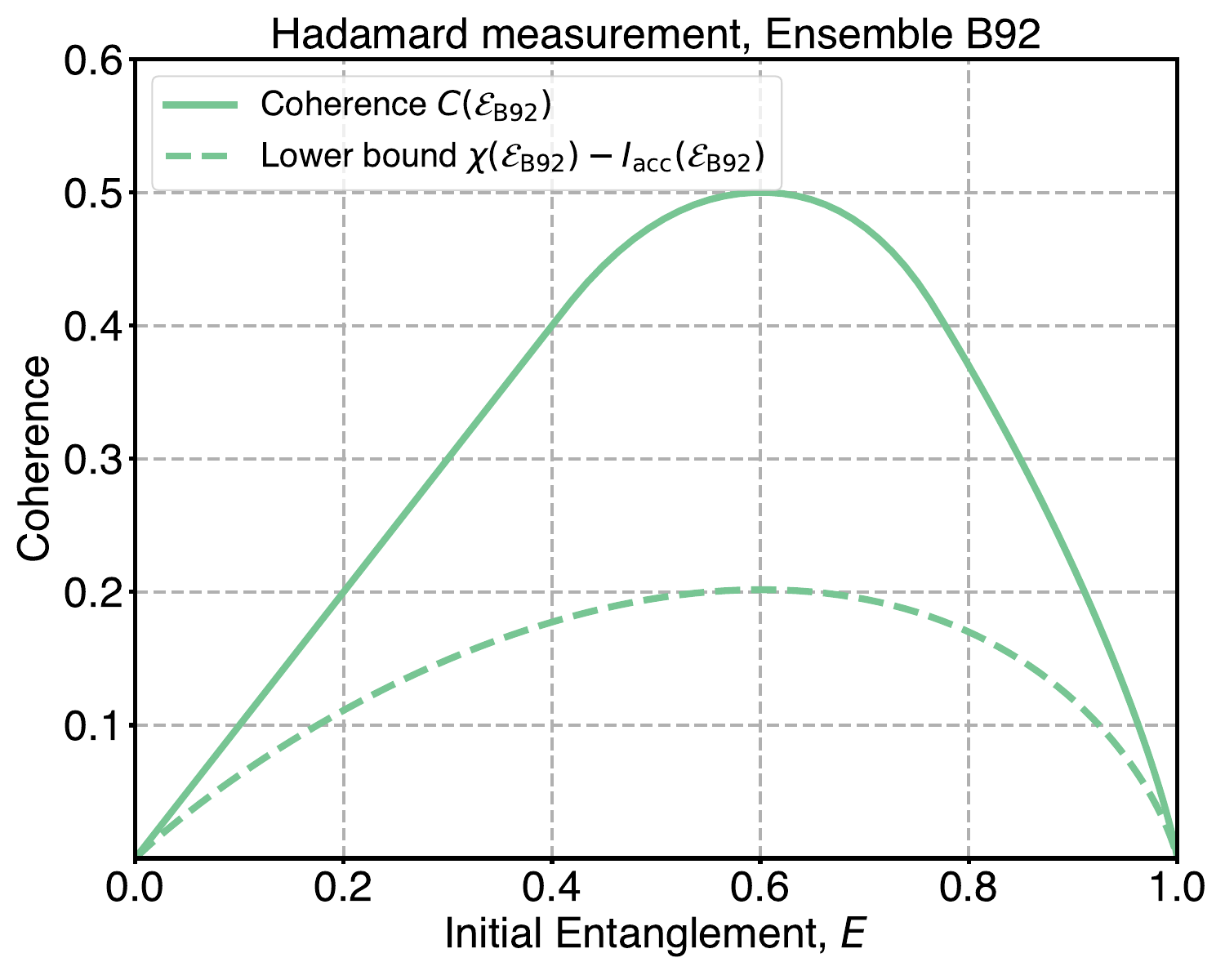}
    \caption{Coherence of two non-orthogonal~states.
    Basis-free coherence and the lower bound on the basis-free coherence as functions of the initial entanglement for a pair of pure non-orthogonal equiprobable states.
    Solid line stands for the coherence values.
    Dashed line corresponds to the values of the lower bound.}
    \label{Fig:Pair_coherence_plot}
\end{figure}

Figure~\ref{Fig:Pair_coherence_plot} shows the minimum coherence value of equiprobable non-orthogonal states as a function of the initial entanglement (solid line).
In the current work, we consider POVM consisting of only rank-1 elements.
Thus, after applying an observable to subsystem A, no residual entanglement is left.
All the initial entanglement is consumed and converted to coherence and accessible information of the ensemble.
As expected above, coherence turns to zero, when the initial entanglement is zero or maximal.
For the small values of the entanglement ($\leq$0.4), the minimum coherence has a linear dependency with the unit slope.
As a result, if~the initial entanglement is less than 0.4, by~applying the Hadamard measurement, we obtain an ensemble with a coherence that is equal to the initial entanglement:
\begin{equation}
    C(\mathcal{E}_\text{B92}) = E, \quad \text{if }E \leq 0.4
\end{equation}
As highlighted in the previous section, the~basis-free coherence of an ensemble is always upper-bounded by its Holevo quantity; see Equation~(\ref{eq:upper}).
For ensembles of pure states, the~Holevo quantity equals the initial entanglement resource.
Within the considered range of initial entanglement, the~coherence reaches this upper bound.
Thus, we identify an optimal strategy within this entanglement range, where coherence attains its maximum achievable~value.

The optimal basis for which minimal coherence value equals entanglement is the standard computational basis $T = \{ \ket{0}, \ket{1} \}$.
In the computational basis, Equation~(\ref{pair_in_basis}) transforms into the entropy of squared Schmidt coefficients, i.e.,~into the initial entanglement measure.
The ensemble is symmetric with respect to the first basis state and tends to align it.
Both density matrices are almost diagonal and, thus, the~coherence naturally reaches its~minimum.

The lower bound on the basis-free coherence is the gap between the Holevo quantity and accessible information of an ensemble and, thus, requires optimization over rank-1 observables.
However, for~the considered ensemble of two pure states, accessible information can be calculated according to the analytical formula~\cite{levitin1995optimal} as well as the Holevo quantity:
\begin{equation}
\begin{gathered}
    I_\text{acc}(\mathcal{E}_\text{B92}) = 1-h_2\left( \frac{1+\sqrt{1-|\langle \psi_1|\psi_2
 \rangle|^2}}{2} \right),\\
    \chi(\mathcal{E}_\text{B92}) = h_2 \left( \frac{1-|\langle \psi_1|\psi_2
 \rangle|}{2} \right).
\end{gathered}
\end{equation}

Figure~\ref{Fig:Pair_coherence_plot} also represents the coherence lower bound obtained analytically as a function of the initial entanglement (dashed line).
The exact values of the lower bound differ from the corresponding coherence values.
Meanwhile, its behavior with the rise in the initial entanglement corresponds well to coherence.
Both quantities reach their maximums when the angle between states of the ensemble is $\pi/4$ or, equivalently, when the initial entanglement $E$ is almost~0.6.

A similar result was obtained in Ref.~\cite{fuchs2002just}, where the advantage of the Holevo quantity over accessible information was proposed as a measure of quantumness for a pair of non-orthogonal states.
Subsequent formal studies on the quantumness of ensembles have also focused on two non-orthogonal states~\cite{luo2009relative, luo_How_quantum, luo2011quantumness, qi2018quantifying, sun2021quantumness, Renyi_2022, fan2024quantifying}.
The ensemble B92 is maximally quantum with respect to most of the quantumness measures when the angle between the states is $\pi/4$.
Refs.~\cite{luo2009relative, luo2011quantumness} provide plots of quantumness measures for the B92 ensemble that are similar to Figure~\ref{Fig:Pair_coherence_plot}.
In this work, we reinterpret the existing results for the B92 ensemble within a more comprehensive entanglement-based framework for ensemble coherence.
In our case, basis-free coherence serves as an upper bound on the gap between the Holevo quantity and accessible information.
Nevertheless, we recover the same conclusion: the ensemble of two non-orthogonal states is most quantum when the angle between them is $\pi/4$.

\section{Symmetric observable}\label{sec:symmetric}

In this section, we present an alternative approach for generating coherence from a given amount of entanglement, based on measurements characterized by high uncertainty.
We introduce a class of symmetric observables with a tunable number of outcomes and adjustable orientation on the Bloch~sphere.

\begin{figure*}[t]
\centering
\includegraphics[scale=0.4]{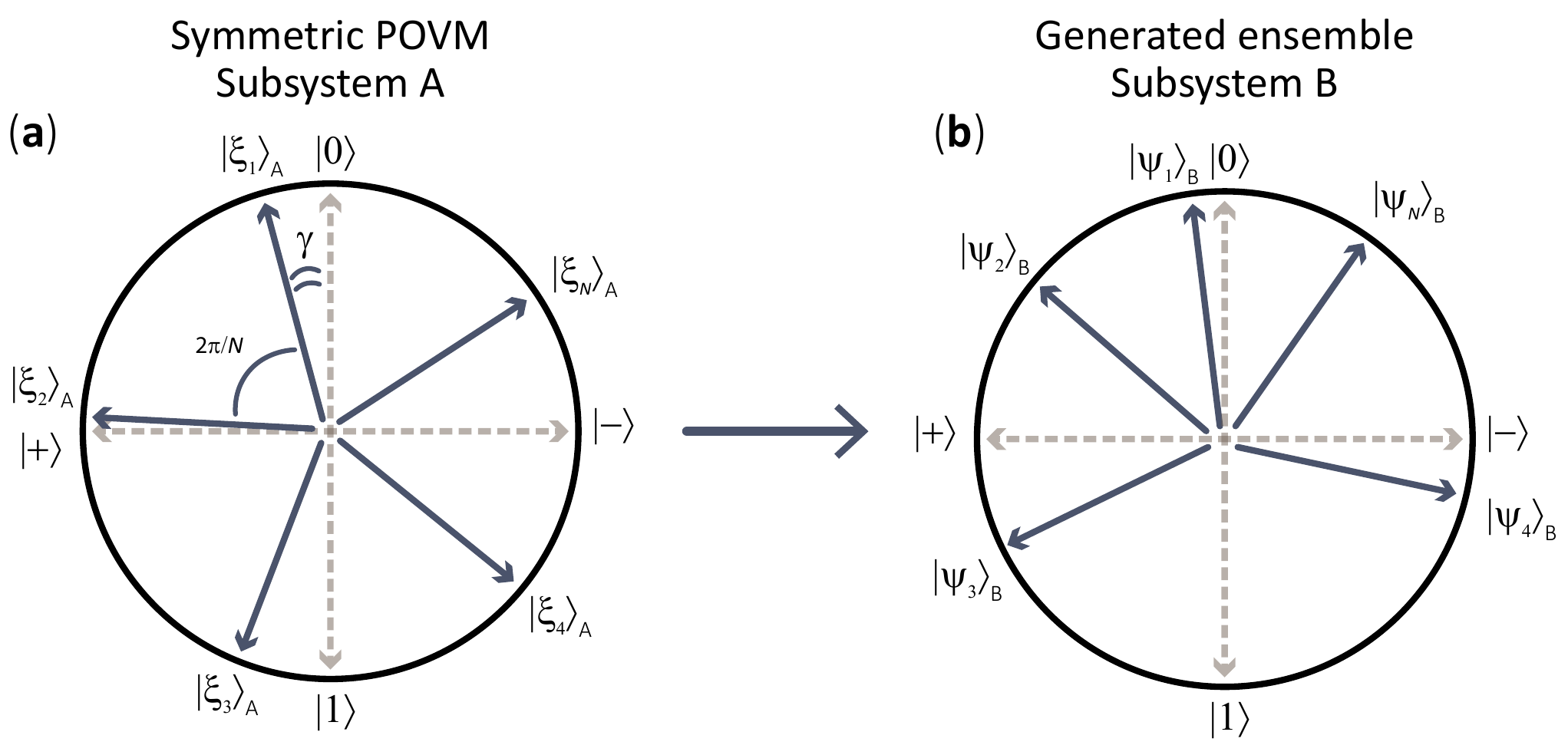}
\caption{Scheme of the ensemble~generation.
    (\textbf{a}) Structure of the rank-1 POVM built upon the set of symmetric states in Alice's subsystem.
    $N=5$.
     (\textbf{b}) Generated ensemble in Bob's subsystem, $N=5$.
    When the initial entangled state is the Bell state, the~ensemble's structure coincides with the symmetric states on Alice's side.
    When the entanglement decreases to zero, Bob's ensemble shrinks to the zero basis state $\ket{0}_\text{B}$.}
\label{fig:Fig2}
\end{figure*}

Specifically, we consider rank-1 POVMs constructed from symmetric pure states~\mbox{\cite{forney1991geometrically, chefles1998optimum, eldar2004optimal, bae2015quantum, lu2023optimal}}, as illustrated in Figure~\ref{fig:Fig2}a.
Increasing the number of POVM elements leads to higher entropy in the measurement outcomes.
The resulting ensemble comprises a large number of quantum states that cannot be simultaneously diagonalized in a common basis.
For each given number of POVM elements, we optimize the orientation of the observable to maximize the resulting coherence.
The central question we address in this section is the following: as the number of POVM elements tends to infinity, what is the limiting value of ensemble coherence?

As in the previous section, we consider an entangled state of qubit subsystems A and B (Equation~(\ref{ent_state})).
Alice applies a local rank-1 observable whose elements are built on the symmetric states:
\begin{equation}
    \mathcal{M}^\text{sym}_N = \left\lbrace \frac{2}{N} \ket{\xi_i}\!\bra{\xi_i} \right\rbrace_{i=1}^{N},
    \label{eq:initial_POVM}
\end{equation}
where
\begin{equation}
\begin{gathered}
    \ket{\xi_i} = \cos\left( \frac{\theta_i}{2} \right) \ket{0}_\text{A}+\sin \left( \frac{\theta_i}{2} \right)\ket{1}_\text{A}, \\
    \theta_i = \gamma + 2 \pi (i-1)/N,
\end{gathered}
    \label{eq:POVM_states}
\end{equation}
where free parameter $\gamma$ is the angle between $\ket{0}_\text{A}$ and $\ket{\xi_1}$, $0 \leq \gamma\leq \pi/N$; see Figure~\ref{fig:Fig2}a.
The states $\ket{\xi_i}$ are symmetric, meaning that each state can be obtained from a previous state by the same unitary transformation $U$: $\ket{\xi_i} = U\ket{\xi_{i-1}}$.
In this case, the~observable is called symmetric with respect to $U$. 
In our case, $U$ has the form
\begin{equation}
    U = \cos \left(\frac{\pi}{N} \right)\mathds{1} - i\sin \left( \frac{\pi}{N} \right) \sigma_y,
\end{equation}
where $\mathds{1}$ is the identity operator and $\sigma_y$ is $Y$ Pauli matrix.
In this section, the~unitary transformation is the amplitude rotation around the $Y$-axis on the Bloch~sphere.

For $N=2$, the~considered observable corresponds to a von Neumann measurement; for $N=3$, it is constructed from the states used in the three-state QKD protocol~\cite{phoenix2000three, boileau2005unconditional}; and for $N=4$, it represents a POVM built from states forming two mutually unbiased bases, similar to those in BB84 QKD~\cite{bennett2014}.
In BB84, to~encode a logical bit, Alice first randomly chooses a basis, either computational or Hadamard.
She then prepares one of two orthogonal states within the chosen basis.
The security of BB84 relies on the fact that it is impossible to correctly identify the state without knowing the initial basis.
A QKD protocol would not be secure if only a single orthogonal basis were used.
A pair of orthogonal states does not exhibit significant quantum properties, i.e.,~they have no coherence, similar to considerations over B92 in Section\,\ref{sec:B92}.
However, when two mutually unbiased bases are merged together within a single ensemble, they cannot be simultaneously diagonalized.
As a result, the~ensemble exhibits coherence, which parallels the secure communication achieved in BB84~QKD.

In the ensemble coherence framework, the~probability of the $i$-th measurement result is
\begin{equation}
    p_i = \frac{2}{N} \left( \cos^2\left( \frac{\theta_i}{2} \right) \cos^2 \alpha + \sin^2\left( \frac{\theta_i}{2} \right) \sin^2 \alpha \right).
\end{equation}
In the case of the maximally entangled Alice--Bob state, $\alpha = \pi /4$, the~resulting probability distribution is~uniform.

The exact form of the obtained ensemble $\mathcal{E}_\text{sym}=\{ p_i, \ket{\psi_i} \}_{i=1}^N$ can be obtained from the operators of the initial POVM and probability distribution of the measurement results:
\begin{equation}
    \ket{\psi_i} = \sqrt{\frac{2}{N p_i}} \left( \cos \alpha \cos\left( \frac{\theta_i}{2} \right) \ket{0}_\text{B} \right.\\
    + \left. \sin \alpha \sin \left( \frac{\theta_i}{2} \right)\ket{1}_\text{B} \right).
\end{equation}

The visual structure of the ensemble's states is presented in Figure~\ref{fig:Fig2}b. 
When the initial entangled state is the Bell state, the~ensemble's structure fully coincides with the states the observable built upon.
In the absence of entanglement, any ensemble fully shrinks into the basis state $\ket{0}_\text{B}$.

To calculate the ensemble coherence, we fix an arbitrary orthogonal basis $T=\{ \ket{e_1},\ket{e_2} \}$ in the subsystem B, where $\ket{e_1} = \cos (\theta_T/2)\ket{0}_\text{B}+\sin (\theta_T/2) \ket{1}_\text{B}$ and $\theta_T$ is a real-value basis parameter.
Coherence in the fixed basis $T$ has the following form:
\begin{equation}
    C_T \left( \mathcal{E}_\text{sym} \right)= \sum\limits_{i=1}^{N} p_i \, h_2 \left( |\braket{e_1|\psi_i}|^2 \right).
    \label{coherence_basis}
\end{equation}
To obtain the coherence minimum over all bases, we have to optimize the latter expression over the basis parameter.
In our optimization, the~basis parameter ranges from 0 to $\pi/2$.
Alice is free to choose the POVM parameter $\gamma$ to maximize the resulting ensemble coherence:
\begin{equation}
    C(\mathcal{E}_\text{sym}) = \max_{\gamma \in [ 0;\frac{\pi}{N} ]} \left( \min_T C_T(\mathcal{E}_\text{sym})\right).
\end{equation}
The parameter $\gamma$ ranges from $0$ to $\pi/N$, as~any further increase does not yield different coherence values due to the symmetry of the considered~observables.

\begin{figure}[t]
\includegraphics[width=9cm]{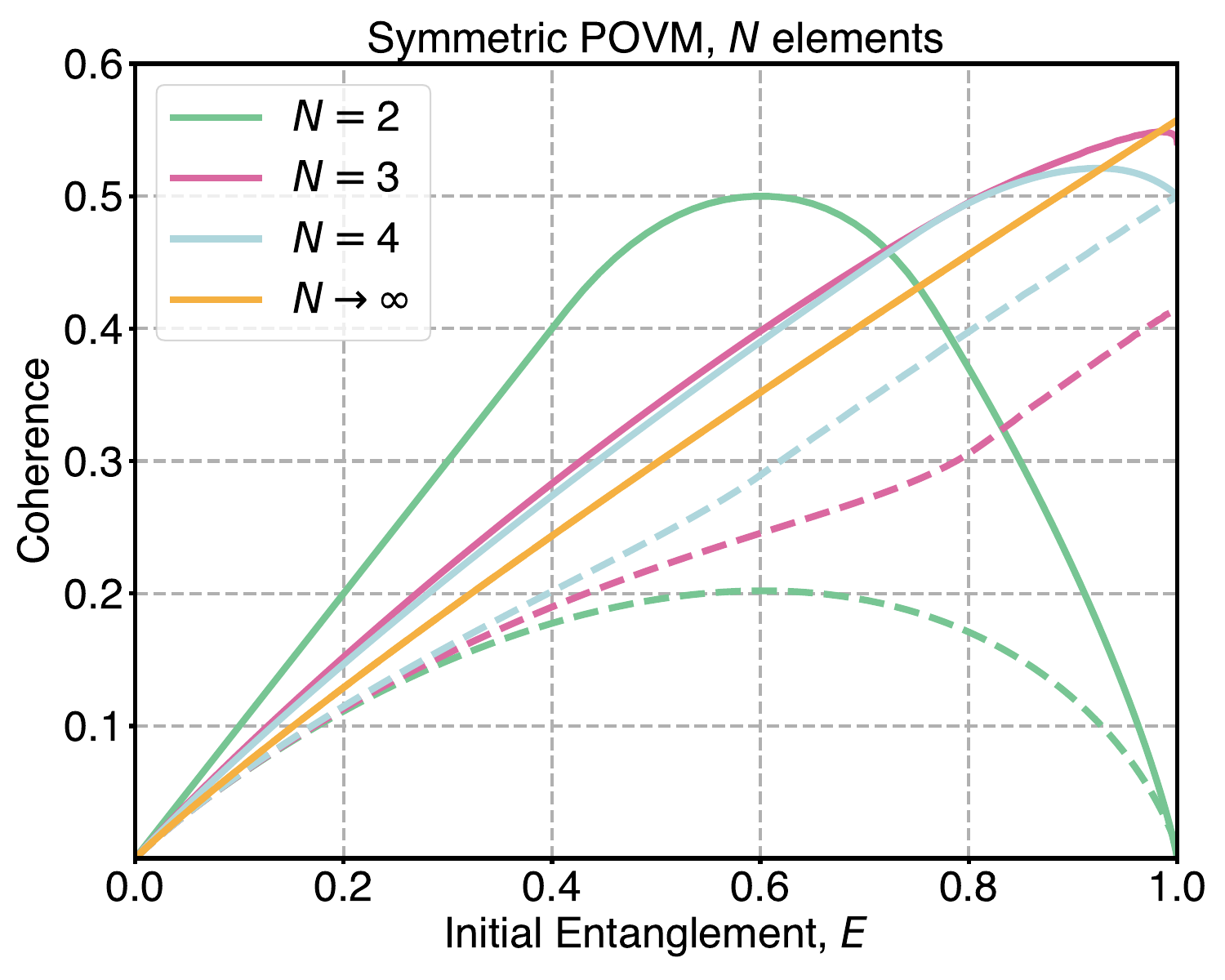}
    \caption{Coherence generated with symmetric~POVM.
    Basis-free coherence and the gap between the Holevo quantity and accessible information as functions of the initial entanglement resource, for~different numbers of POVM elements $N=2,\,3,\,4.$
    The line labeled $N\rightarrow \infty$ corresponds to the asymptotic behavior.
    Different colors correspond to different values of $N$.
    Solid lines represent coherence values, while dashed lines indicate the corresponding lower bounds.
    For $N\rightarrow \infty$, the~solid and dashed lines coincide.}
    \label{fig:Coherence_Iacc}
\end{figure}

Figure~\ref{fig:Coherence_Iacc} shows the dependence of the ensemble's informational characteristics on the initial entanglement for different numbers of elements in the applied observable.
The solid lines correspond to the basis-free coherence.
Meanwhile, dashed lines represent the lower bound on coherence, based on the gap between the Holevo quantity and the accessible information.
Colors of the lines reflect the number of elements in the POVM.
The case of $N=2$ exactly matches the data shown in Figure~\ref{Fig:Pair_coherence_plot}.
The optimal rotation angle of the initial observable, $\gamma$, strictly depends on the number of POVM elements.
For an even $N$, the~optimal rotation angle is $\pi / N$ and is independent of entanglement.
For an odd $N$, the~optimal $\gamma$ equals $\pi / N$ for the maximally entangled state and then decreases monotonically toward zero as entanglement decreases.
Figure~\ref{fig:Coherence_Iacc} also shows the dependence of coherence on entanglement for an asymptotically large number of POVM elements, labeled as $N \rightarrow \infty$ (orange line).
The coherence values rapidly reach saturation as $N$ increases.
In this regime, the~maximum coherence is about $0.56$ for $N \rightarrow \infty$ and is naturally achieved for a maximally entangled~state.

The difference between the basis-free coherence of an ensemble and the lower bound, defined through the Holevo quantity and accessible information, decreases as $N$ increases.
For $N = 2$, the~maximal gap between coherence and the lower bound is $0.3$, while for $N = 4$ the maximal gap is $0.1$.
According to numerical data, in~the asymptotic limit of the number of POVM elements, the~maximal gap is only $10^{-8}$, which corresponds to the accuracy of the numerical optimization.
As a result, there is no dashed line for $N \rightarrow \infty$ in Figure~\ref{fig:Coherence_Iacc}, since it practically coincides with the solid line representing coherence within numerical accuracy.
Thus, numerical data indicate that, in~the limit of sufficiently large $N$, the~lower bound approaches the coherence, and~the following equality holds:
\begin{equation}
    E = \lim_{N\rightarrow \infty} \left( C(\mathcal{E}_\text{sym}) + I_\text{acc}(\mathcal{E}_\text{sym}) \right).
\end{equation}
The initial entanglement splits into the ensemble's coherence and accessible~information.

Depending on the amount of entanglement, different observables yield maximal coherence.
Within the considered class of POVMs, the~two-element observable is optimal for most values of entanglement.
In the vicinity of maximal entanglement, the~optimal POVM corresponds to a sufficiently large number of elements $N$.
As a result, no single POVM within the class can be identified as universally optimal for coherence generation.
In contrast, the~behavior of the gap between the Holevo quantity and accessible information exhibits a different trend: this gap increases with increasing $N$.
For example, the~observable with three elements consistently yields a higher lower bound value compared to the case of two non-orthogonal states.
Similarly, a~BB84-like observable outperforms POVM with $N=3$ in terms of the gap between the Holevo quantity and accessible information; see dashed lines in Figure~\ref{fig:Coherence_Iacc}.

Similar ensembles have been considered as specific examples in several studies~\cite{fan2024quantifying, luo2011quantumness, Renyi_2022, sun2021quantumness}.
In this work, we reinterpret these ensembles within an entanglement-based framework using the class of symmetric observables.
The ensembles studied in the aforementioned works correspond to performing corresponding measurements on one half of a maximally entangled state.
For instance, to~generate the BB84 ensemble on Bob's side, one must apply the BB84 POVM on Alice's subsystem.
Our approach reveals how the properties of the resulting ensemble depend on the amount of initial entanglement resource.
While Refs.~\cite{fan2024quantifying, Renyi_2022, sun2021quantumness} report a strict ordering of quantumness among ensembles, our entanglement-based analysis shows that the optimality of a given ensemble depends on the available entanglement, as~discussed~above.

\bigskip
\section{Conclusion}
In this study, for~the qubit case, we demonstrate how an ensemble with explicit quantum properties~\cite{kronberg2019coherence, kodukhov2023entanglement} can be generated via a von Neumann measurement in the Hadamard basis, which is unbiased with respect to the computational basis forming the Schmidt basis.
The resulting ensemble consists of non-orthogonal states corresponding to the B92 QKD protocol~\cite{bennett1992quantum}; see Section~\ref{sec:B92}.
This ensemble exhibits zero coherence when the initial entanglement is either zero or maximal~\cite{bennett1996concentrating, bennett1996purification, wootters1998entanglement}.
This behavior has a clear interpretation in the context of quantum cryptography:
When the initial state is maximally entangled, the~resulting ensemble consists of orthogonal states which can be perfectly distinguished by an eavesdropper without being detected.
Conversely, in~the absence of entanglement, the~ensemble collapses to identical states, making it impossible for the receiver to distinguish logical bits.
Both extremes result in a zero secret key generation rate.
The considered pair of non-orthogonal states is maximally quantum when their overlap is $1/\sqrt{2}$---a result consistent with the findings initiated by C.A.\,Fuchs~\cite{fuchs2002just} and developed further in a number of studies~\cite{luo2009relative, luo_How_quantum, luo2011quantumness, qi2018quantifying, sun2021quantumness, Renyi_2022, fan2024quantifying}.
Notably, the~basis-free coherence of the ensemble equals the initial entanglement resource for entanglement values up to $E \leq 0.4$.
Within a specific range of entanglement, the~coherence reaches its upper bound, indicating that the proposed strategy for entanglement-to-coherence conversion is~optimal.

As a second example, we consider a class of symmetric observables; see Section~\ref{sec:symmetric}.
Each observable in this class is a rank-1 POVM constructed from symmetric states~\cite{forney1991geometrically,chefles1998optimum, eldar2004optimal, bae2015quantum, lu2023optimal}.
In the case of maximal entanglement, this class allows the generation of ensembles corresponding to BB84~\cite{bennett2014} and three-state QKD~\cite{phoenix2000three, boileau2005unconditional} protocols.
The number of POVM elements can be increased indefinitely, leading to a saturation point.
In this regime, the~initial entanglement equals the sum of the accessible information~\cite{divincenzo2004locking, konig_renner_2007, pastushenko2023improving} and the basis-free coherence~\cite{kodukhov2023entanglement} of the ensemble.
Depending on the given amount of entanglement, different ensembles exhibit varying degrees of coherence.
In contrast to the ordering of ensemble quantumness found in previous studies~\cite{fan2024quantifying, Renyi_2022, sun2021quantumness}, we observe that the optimality of a particular ensemble depends on the amount of entanglement.
Generally, computing ensemble coherence requires knowing the exact structure of the states in the ensemble after Alice's measurement.
However, in~the present case the considered class of POVMs consists only of rank-1 observables, and~therefore the joint Alice–Bob Born-rule probabilities are sufficient to calculate the required entropies; see Equation~(\ref{eq:rel_ent}).

In principle, we can apply a general, non-rank-1 observable.
In this case, the~obtained ensemble would include mixed states.
Thus, in~the coherence definition (Equation~(\ref{eq:rel_ent})), the~second term would play an essential role.
To calculate the ensemble coherence, it would not be enough to have only the Alice–Bob Born-rule probabilities; we would also need the exact form of Bob's states.
Since the ensemble would contain mixed states, there would be a non-zero residual entanglement between Alice and Bob~\cite{bennett1996concentrating, bennett1996purification, wootters1998entanglement}.
As a result, such an observable would produce less ensemble coherence, since not all of the initial entanglement resource would be~consumed.

Within the class of observables considered in this work, we did not identify a universal strategy for obtaining coherence equal to the initial entanglement.
Further strategies for constructing such a class of observables will require extending the analysis to the entire Bloch sphere, rather than restricting it to the $XZ$-plane.
This extended class should include, for~example, the~six-state QKD~\cite{bechmann1999incoherent} observable and the symmetric, informationally complete POVM (SIC-POVM).

In addition, this work focuses exclusively on pure bipartite entangled states.
In the case of more than two entangled systems, a~measurement on one subsystem produces ensembles in the remaining subsystems.
The scenario of three entangled systems is of special interest from the perspective of quantum cryptography.
The most well-known security approach, the~Shor--Preskill security proof, considers an entangled state shared among Alice, Bob, and~an eavesdropper who controls the entire environment~\cite{shor2000simple}.
Extending the concept of ensemble coherence to the framework of the Shor--Preskill security proof will be the subject of future~research.

\bibliography{re}
\end{document}